\documentclass[amsmath,amssymb,nofootinbib,aps]{revtex4}
\usepackage{graphicx}
\usepackage{dcolumn}
\usepackage{color}
\usepackage{bm}
\usepackage{amsmath,amssymb,graphicx,hyperref,subfigure,multirow,subcaption}
\usepackage{epsfig}
\usepackage{enumerate}
\usepackage{array}
\usepackage{multirow}

\newcommand{\sech}{\textrm{sech}}

\begin{document}
\title
{Accessing Levinson-Smith and Li\'{e}nard equations in a nonlinear framework of quantum mechanics}

\author{Bijan Bagchi,\footnote{E-mail: bbagchi123@gmail.com}$^1$, 
 Anindya Ghose-Choudhury\footnote{E-mail: aghosechoudhury@gmail.com}$^2$}

\vspace{2mm}

\affiliation{$^{1}$Department of Applied Mathematics, University of Calcutta, Kolkata 700009, India\\
$^2$Department of Physics, Diamond Harbour Women’s University,
Diamond Harbor Road, Sarisha, West Bengal 743368, India}

\vskip-2.8cm
\date{\today}
\vskip-0.9cm

\begin{abstract}

In this paper, we present the reduction of a recently introduced generalized quantum scheme of Chodos and Cooper to Levinson-Smith and  Li\'{e}nard classes of nonlinear system that enable us to furnish a wider class of acceptable solutions. In particular, for the Levinson-Smith equations, we find their relevance to position-dependent mass quantum systems, with an interesting off-shoot that solitonic-like solutions emerge from the condition of the level surface in the system. On the other hand, for the Li\'{e}nard type, which has coefficients of odd and odd symmetry in the present set up, we demonstrate that closed form solutions exist on carrying out conversion to the typical Abel form.  \\

\end{abstract}

\maketitle

 \textbf{Keywords:} Nonlinear quantum mechanics, Li\'{e}nard equation, Jacobi Last Multiplier, exact solutions \\\\
 
 \textbf{PACS numbers:} 3.65Ud, 3.657a, 05.45.-a, 11.10.Lm

\section{Introduction}

More than three decades and a half ago, Weinberg \cite{weinberg1, weinberg2} made an attempt to carry out possible tests of nonlinearity while developing plausible generalizations of quantum mechanics, 
and exploring equations 
which were of Hamiltonian type. In an effort to set stringent limits on possible nonlinear corrections, he introduced detuning of resonant transitions. Apart from nonlinearities leading to a violation of the second law of thermodynamics \cite{peres}, such a model raised some important questions in connection with instantaneous communication between different quantum states \cite{polchinski, gisin1, gisin2}. 
In fact, it was also realized that Weinberg's work may not be free from pathological problems that could allow superluminal communication through entangled Einstein-Podolsky-Rosen pairs, and in the process, violated relativistic causality (see also \cite{czachor1, czachor2}). Of course, contrary view was held \cite{mielnik} which pointed out that if any enlargement of a structural scheme is carried out in which the conventional notion of probability for pure and mixed states is held, then this would inevitably trigger off causality violation. We should mention that even earlier there has been a body of work devoted to stochastic, nonlinear modification of the Schr\"{o}dinger equation \cite{ghi}. In addition, new operator integrability conditions have been exploited \cite{hsu}, while using certain postulates, several fundamental aspects have been examined in which the
implicit wave function of linear quantum mechanics was shown to be determined by solving a nonlinear partial differential equation \cite{row}. In the literature, the compatibility of nonlinear quantum mechanics with artificial intelligence has also been a topic of attention \cite{opp}.\\

In more recent times, a nonlinear heuristic scheme of quantum mechanics (NLQM) was proposed in a series of papers by Chodos and Cooper \cite{cc1, cc2, cc3}. Their motivation to undertake such a study came partly in a bid to examine some of the viable consequences that might prove to be of academic interest, but that could also allow one to examine how gravity went with quantum mechanics, in schemes other than those which are already known like string theory or loop quantum gravity. In this regard, in view of the earlier studies on certain effective quantum schemes where interactions were introduced to characterize nonlinear behavior, they took up a simple nonlinear manifestation of conventional quantum mechanics involving a connection between two fundamental state vectors. The basic idea was to construct a model \cite{cc1, cc2}, in parallel with the well known pair of nonlinear Schr\"{o}dinger equations, that could be described in terms of a pair of underlying ket vectors $|\psi\rangle$ and $|\phi\rangle$, with an additional $\langle \phi|\psi\rangle$ term accounting for the nonlinear effects. In a later work, the model was generalized \cite{cc3} whose Hamiltonian formulation was described by a coupled set of first-order differential equations. \\

Our purpose in this note is to point out that the basic equations residing in the minimal version of the NLQM scheme can be profitably exploited to track down certain well known nonlinear systems.
 In particular, we show that connections to 
Levinson-Smith \cite{lev1, lev2} class of second-order ordinary differential equations (ODEs), a particular case of which contains the Li\'{e}nard form \cite{lie1, lie2}, can be readily established. By tuning the control parameters, we also provide viable solutions for them. A few words on these equations are in order. While Levinson-Smith type of ODE describes self-sustained oscillations along with relaxation oscillations, and yields to novel Liouvillian integrable subfamilies \cite{demi}, from the application perspective, Li\'{e}nard equation appears in many branches of physics such as in optics \cite{dor}, shallow water-wave studies \cite{zho} and non-Hermitian quantum mechanics \cite{bag1}, as well as at other places \cite{pgag, agpgk}. Furthermore, in certain cases, its governing Hamiltonian through the employment of Jacobi Last Multiplier (JLM) \cite{whitt} straightforwardly gives the associated Lagrangian \cite{car}. \\

The paper is organized as follows. In Section II we give a brief summary of the NLQM model highlighting its salient features. In Section III we take up connections to Levinson-Smith and Li\'{e}nard classes of equations. Specifically, Section II A addresses the Levinson-Smith type of equations where we point out relevance to position-dependent mass (PDM) quantum systems, and note extraction of different types of solitonic-like solutions from the level surface condition. In Section III B we consider the class of Li\'{e}nard equations where we show first by transformation to the Abel form and then by an appropriate change of variable how reduction to a first-order form is possible that is immediately solvable. In Section IV some concluding remarks are given.\\

\section{The NLQM model}

To develop a minimal class of NLQM, one replaces the standard Schr\"{o}dinger equation\footnote{The unit $\hbar = 1$ is adopted.} $i\frac{\partial}{\partial t}|\psi \rangle = H |\psi \rangle $, defined in terms of a single ket-vector $|\psi \rangle$ and described by the time-dependent Hamiltonian $H$, by two different representations which are guided by the ket vectors $|\psi \rangle$ and $|\phi \rangle$ which are intertwined as reads below 

\begin{eqnarray}
&& i\frac{\partial}{\partial t}|\psi \rangle = H |\psi \rangle + g |\phi \rangle  \langle \phi|\psi \rangle \label{nlqm1}\\
&& i\frac{\partial}{\partial t}|\phi \rangle = H |\phi \rangle + g^*  \langle \psi|\phi \rangle |\psi \rangle \label{nlqm2}
\end{eqnarray}
In (\ref{nlqm1}) and (\ref{nlqm2}), the interaction coupling may be complex: $g \in \mathbb{C}$. Explicitly, we take $g = a + ib$, where $b \neq 0$. \\ 

In matrix notation, the above 2-parameter formulation can be cast as a single equation

\begin{equation}\label{nlqm3}
    i\frac{\partial}{\partial t}|\Psi \rangle = \mathcal{H} |\Psi \rangle + \mathcal{S}|\Psi \rangle 
\end{equation}
where $|\Psi \rangle = \left (|\psi \rangle, |\phi \rangle \right )^T$ and $\mathcal{S}$ represent the Hermitian matrix

\begin{equation} \label{nlqm4}
\mathcal {S} = \begin{bmatrix} 0  & g\langle \phi|\psi \rangle \\ g^*\langle \psi|\phi \rangle & 0  \end{bmatrix}  
\end{equation}\\
The fundamental quantities which play a crucial role in the dynamics of the system are given by 

\begin{equation}
N = \langle \psi|\psi \rangle + \langle \phi|\phi \rangle, \quad y = \langle \psi|\psi \rangle - \langle \phi|\phi \rangle, \quad \gamma =  \langle \phi|\psi \rangle, \quad x = |\gamma|^2 
\end{equation}\\
It may be remarked that under phase transformations, the equations of motion remain invariant. Their time dependence can be worked out straightforwardly. It turns out that

\begin{equation}\label{nlqm5}
\frac{dN}{dt} = 0, \quad \frac{d y}{dt} = 4bx, \quad \frac{d\gamma}{dt} = ig \gamma y
\end{equation}\\
Note that, to this end, the Hermiticity of $\mathcal {S}$ proves to be a sufficient criterion: the constancy of the norm ensures that the evolution of the state vector is bounded despite the presence of the complex character of $g$. The above equations can be comprehensively solved to generate specific classes of solutions that are of potential physical interest. We do not go into the details of these calculations which are available in \cite{cc3}.\\

Technically, the Hermiticity of $\mathcal {S}$ although sufficient is not a necessary requirement. It can have an anti-Hermitian part as given by the following matrix structure 
 
 \begin{equation}\label{nlqm6}
\mathcal {T} = \begin{bmatrix} i\mu\langle  \phi|\phi \rangle  & \lambda\langle \phi|\psi \rangle \\ -\lambda\langle \psi|\phi \rangle & -i\mu\langle \psi|\psi \rangle  \end{bmatrix}  
\end{equation}\\
where $\mu \in \Re$, leading to a very general form provided by the combination $\mathcal{S} = \mathcal{\tilde{S}} + \mathcal{T}$, where 

\begin{equation}\label{nlqm7}
\mathcal{\tilde{S}} = \begin{bmatrix} \alpha_1 \langle \psi|\psi \rangle + \alpha_2 \langle \phi|\phi \rangle & g\langle \phi|\psi \rangle \\ g^*\langle \psi|\phi \rangle & \beta_1 \langle \psi|\psi \rangle + \beta_2 \langle \phi|\phi \rangle\end{bmatrix}  
\end{equation}\\
Note that $\alpha_i, \beta_i \in \Re$ (i=1, 2) and that $\mathcal{\tilde{S}} = \mathcal{\tilde{S}}^\dagger$. In \cite{cc3}, such a scheme was exploited to implement a natural extension of the Born rule for the computation of particle trajectories and observing that closed particle orbits are possible.\\


In such a general setting, the variables $x$ and $y$ can be straightforwardly seen to obey the following system of coupled nonlinear differential equations

\begin{eqnarray}
&& \frac{d x}{dt} = -2\left (b + \mu \right )yx \label{nlqm8}\\
&& \frac{d y}{dt} = \mu \left (N^2 - y^2\right ) + 4 b x \label{nlqm9}
\end{eqnarray}
Since closed-form exact solutions are hard to find, the integrability of (\ref{nlqm8}) and (\ref{nlqm9}) can be examined for some special cases of the parameters $\mu$ and $b$. In \cite{cc3}, the hyperbolic solutions for $y$ and $z$ were found to be preferred over non-physical trigonometric ones. In particular, the simpler case of $y = 0$ (or $b = 0$) pointed to $y$ and $x$ emerging as a combination of $\tanh t$ and $\cosh^{-2} t$, respectively.  \\

We also see that a simplification of equation (\ref{nlqm9}) takes place by fixing the parameter $b=-2\mu$ which gets rid of the damping term. As a result, on eliminating $x$, we find

\begin{equation}\label{nlqm13}
\frac{d^2y}{dt^2}+2\mu^2 N^2y-2\mu^2 y^3=0
\end{equation}
An inspection with the Jacobi elliptic function $sn(u, k )$ equation which reads

\begin{equation}\label{nlqm14}
\frac{d^2w}{du^2}+(1+k^2)w-2k^2 w^3=0, \quad 0<k<1
\end{equation}
immediately shows that $y(t)=sn(t, \mu)$ solves (\ref{nlqm13}) subject to the constraint, $N^2=\frac{\mu^2+1}{2\mu^2}$, where $0<\mu<1$. It then transpires from (\ref{nlqm8})

\begin{equation}\label{nlqm15}
x(t)=\frac{\mu^2+1}{16\mu^2}-\frac{1}{8}sn^2(t, \mu)-\frac{1}{8\mu}cn(t, \mu)dn(t, \mu)
\end{equation}
is a viable solution for it. \\

Some remarks are in order on the interpretation of (\ref{nlqm8}) and (\ref{nlqm9}) as constituting a dynamical system. We observe that the points (i) $x=0, \ y= \pm N$, and (ii) $x= - \frac{\mu N^2}{4b} , \ y = 0$ are plausible candidates for equilibrium points. The Jacobian matrix which is given by 

\begin{equation}\label{nlqm10}
\mathcal {J} = \begin{bmatrix} \mp2(b+\mu)N  & 4b \\ 0 & \mp2\mu N  \end{bmatrix}  
\end{equation}\\
yields the characteristic equation 

\begin{equation}
\lambda^2 -(tr J) \lambda + \det J = 0 
\end{equation}\\
in which 

\begin{equation}
tr J = \pm 2N(b+2\mu)
\end{equation}\\
along with 

\begin{equation}
\det \ J = 4N^2\mu (b+\mu)
\end{equation}\\
Using Routh-Hurwitz's criterion of stability \cite{sch, laks}, it turns out that for $\mu > 0$ and $b+\mu > 0$, the set of coordinates $x=0, \ y=N$ is a stable point, while $x=0, \ y= - N$ is unstable for these conditions. The other equilibrium point $(-\frac{\mu N^2}{4b}, 0)$ is of the saddle type and is therefore unstable for all the parametric regions.\\




\section{Connecting to Levinson-Smith and Li\'{e}nard equations}

To tackle a more general situation, we take a derivative of equation (\ref{nlqm8}) and use (\ref{nlqm9}) to eliminate $y$. As a consequence, $x (t)$ is found to satisfy the following second-order differential equation

\begin{equation}\label{nlqm11}
\frac{d^2 x}{d t^2} - \frac{(2b + 3\mu)}{2(b+\mu)} \frac{1}{x} \left (\frac{d x}{dt}\right)^2 + (\mu + b)\left [2\mu N^2 x + 8b x^2 \right ] = 0
\end{equation}
Similarly, elimination of $x$ shows that $y (t)$ obeys

\begin{equation}\label{nlqm12}
\frac{d^2 y}{d t^2} + 2(2\mu + b)y \frac{d y}{dt} + 2\mu (\mu + b)y(y^2 - N^2) = 0
\end{equation}
where we assume for the present $b+\mu \neq 0$. These are the basic ODEs which we would explore below. A word about their classification is in order. The ODE corresponding to the $x$-variable corresponds to the Levinson-Smith class \cite{lev1, lev2}. It sets the structural foundation, after which the ODE corresponding to the $y$-variable, following a logical specialization, assumes the Li\'{e}nard form on enforcing transformation to the Abel's form.\\

\subsection{Levinson-Smith type}

The Levinson-Smith family of ODEs \cite{lev1, lev2} is given by the form 

\begin{equation}\label{nlqm22}
\frac{d^2 s}{d t^2} + J (s) \left (\frac{ds}{dt}\right )^2 + F (s) \left (\frac{ds}{dt}\right ) + G(s) =0 
\end{equation}
where $s \in \Re$, and $J(s),\ F(s)$ and $G(s)$ are differentiable functions of $s$. In the present case of (\ref{nlqm11}), $F=0$, and the quantities $J$ and $G$ correspond to the functions

\begin{equation}\label{nlqm23}
J = -\frac{(2b + 3\mu)}{2(b+\mu)} \frac{1}{x}, \quad   G  = (\mu + b)\left [2\mu N^2 x + 8b x^2 \right ]  
\end{equation}

One can express $J$ in the form of a total derivative, i.e., 

\begin{equation}
J = \frac{d(\ln(x)^{-\Lambda})}{d x}, \quad \Lambda = \frac{(2b + 3\mu)}{2(b+\mu)}
\end{equation}\\
Therefore, multiplying (\ref{nlqm11}) by the quantity $\exp \left(-2\Lambda \int \frac{d x}{x} \right)$, which is basically the JLM \cite{nucci}, and represents the factor $x^{-2\Lambda}$, its first integral projects out as

\begin{equation}\label{nlqm24}
\mathcal{E}\left (x, \frac{dx}{dt} \right ) = \frac{1}{2}x^{-2\Lambda} \left (\frac{d x}{dt} \right)^2 - 2(\mu + b)^2 \left [N^2 x^{-\frac{\mu}{\mu+b}} - 4 x^{\frac{b}{\mu+b}} \right ]
\end{equation}
The above constraint is indicative that a PDM system is at work. Written explicitly, the mass variable $\mathcal{M}(x)$ and the potential energy $V(x)$ read \\

\begin{eqnarray}
&& \mathcal{M}(x) = x^{-2\Lambda} \label{nlqm25}\\
&& V(x) = - 2(\mu + b)^2 \left [N^2 x^{-\frac{\mu}{\mu+b}} - 4 x^{\frac{b}{\mu+b}} \right ] \label{nlqm26}
\end{eqnarray}\\
Interpreted in the framework of quantum mechanics, because of the momentum and PDM noncommutativity, detailed
operator ordering is required for the kinetic part of the Hamiltonian to ensure Hermiticity and overall consistency
of the system\footnote{However, from a classical perspective, the PDM behavior would disappear completely through a change of variable, thus depicting a particle of constant mass in a potential made of power-law terms.}. It may not be out of place to remark that the frequently used effective-mass kinetic energy operator in one dimension was given by von Roos \cite{von}, which in two-parameter form reads 

\begin{equation}
\hat{T} = \frac{1}{4} \left [m^\alpha (x) \hat{p}m^\beta (x) \hat{p}m^\gamma (x) (x) + m^\gamma (x) \hat{p}m^\beta (x) \hat{p}m^\alpha (x) (x) \right ]
\end{equation}\\
where $\hat{p}$ stands for the momentum operator, $m(x)$ represents the position-dependent mass, while the parameters $\alpha, \beta, \gamma$ are subjected to the constraint $\alpha + \beta +\gamma = -1$. Since the momentum operator and position-dependent mass do not commute,
maintaining Hermiticity of $\hat{T}$ is not trivial. Here it is specifically designed to be Hermitian.
Although initially pursued in condensed matter physics problems (see, for example, \cite{bast}) the field of PDM has flourished in many directions over the years (see \cite{ques1, ques2}). Indeed, varying mass problems have provided useful insights into new classes of phenomena that are hard to reconcile with the standard constant-mass prescription. Some of the physical problems in which the idea of PDM has been implemented include those of compositionally graded crystals \cite{geller}, quantum dots \cite{serra}, and liquid crystal problems \cite{barran}.  \\

In (\ref{nlqm25}), for $\Lambda > 0$, potential singularity appears at x = 0, and the usual prescription requires 
to split the real axis into two halves, i.e., $x > 0$ and $x < 0$, where the dynamics of the problem becomes exactly solvable for both branches. The singular  mass profile has also been studied in the context of a one-dimensional crystalline lattice \cite{lima}. Concerning the potential function $V(x)$, we see that it can also have a singular behavior at $x=0$ depending on the choice and sign of the parameters. The appearance of singular potentials is common in quantum mechanics: for instance, in the appearance of negative energy states and the encountering of the degeneracy problem \cite{panigrahi}. On the other hand, the mass profile includes a power-law behavior for $\Lambda < 0$. In the literature, power representations for PDM have been studied for different contexts and their spectral signatures have been investigated \cite{mustafa}. \\

On a level surface $\mathcal{E}(x, \frac{dx}{dt}) = E$, where we assume $E > 0$, the following restriction holds
\begin{equation}\label{nlqm27}
\frac{1}{2}x^{-2\Lambda} \left (\frac{d x}{dt} \right)^2  = E +  2(\mu + b)^2 \left [N^2 x^{-\frac{\mu}{\mu+b}} - 4 x^{\frac{b}{\mu+b}} \right ]
\end{equation}
which can be transformed to the integral 

\begin{equation}\label{nlqm28}
t_0 \pm t = \int \frac{d x}{\left [2Ex^{\frac{3\mu +2b}{\mu+b}} + 4(\mu + b)^2 N^2 x^2 - 16 (\mu + b)^2 x^3 \right ]^{\frac{1}{2}}}
\end{equation}
where $t_0$ is an initial value of time.\\

To generate solutions that are tractable mathematically and which yield solitonic-like profile, let us focus on the simple case $b = 0$. Then the following form is implied

\begin{equation}\label{nlqm29}
t =- \int \frac{d x}{\left [4\mu^2 N^2 x^2  +(2E - 16 \mu^2)x^3 \right ]^{\frac{1}{2}}}, \quad \mu \neq 0
\end{equation}
where we have ignored the constant of integration and applied the negative sign to ensure the positivity of both $E$ and $x$.  The above form corresponds to the standard type of solvable integral $\int \frac{dw}{w(1\pm aw)^{\frac{1}{2}}}$, with $a$ standing for $a=\frac{2E-16\mu^2}{4\mu^2 N^2}$, and where the two signs depend on whether $E > 8 \mu^2$ or $E < 8 \mu^2$. \\
 
Solving (\ref{nlqm29}) we get the expression 

\begin{equation}\label{nlqm30}
    x (t) = \left (\frac{2\mu^2N^2}{8\mu^2 - E} \right) \sech^2 (N\mu t), \quad E < 8\mu^2
\end{equation}
which can also be looked upon as a solution that emerges as a special case from (8), (9), (14) and (15). Since $x$ stands for the positive-definite quantity $|\gamma|^2$, which incidentally corresponds to the square of the amplitude, it can be regarded as a kind of probability density. Thus, the solution becomes regularized for large values of $E$. It shows the existence of the well known uniformly bounded solitonic profile in which the argument is represented by a scaled variable of $t$. The above solution is in tune with the criterion of stability, noted earlier in the context of interpreting (\ref{nlqm8}) and (\ref{nlqm9}) as representing a dynamical system for the conditions $\mu > 0$ and $b+\mu > 0$.\\

Once the exact profile for $x(t)$ is known, $y(t)$ is directly determined by algebraically rewriting equation (9) in the form

\begin{equation}
y(t) = -\frac{1}{2\mu} \frac{d}{dt} \ln x(t), \quad b=0
\end{equation}
For instance, using the above solitonic profile in the form $x(t) = A \operatorname{sech}^2(\Omega t)$, the solution $y(t)$ corresponds to a smooth, bounded, anti-symmetric profile 

\begin{equation}\nonumber
y(t) = \frac{\Omega}{\mu} \tanh(\Omega t)
\end{equation}
This clearly shows that as $t \to \pm\infty$, $y(t)$ asymptotes to stable boundary values $\pm \frac{\Omega}{\mu}$, representing a smooth dynamical population transfer between the state vectors $\vert{}\psi\rangle$ and $\vert{}\phi\rangle$. \\

The transposed case where $\mu = 0$ but $b \neq 0$ is also similar. Discarding the constant of integration, we are led to the integral 

\begin{equation}\nonumber
t = -\int \frac{dx}{\left [(2E + 4b^2 N^2 )x^2  -16 b^2 x^3 \right ]^{\frac{1}{2}}}, \quad b \neq 0
\end{equation}
where again the negative sign is kept. It belongs to the type $\int \frac{dw}{w(h-w)^{\frac{1}{2}}}$, with $h$ given by $h=\frac{2E+4b^2 N^2}{16 b^2}$. Since the parameter $h$ here is intrinsically positive for $E > 0$, the solution for $x(t)$ can be easily extracted in the form


\begin{equation}\label{nlqm32}
x(t) = \frac{1}{8b^2}\left (E + 2 b^2 N^2 \right ) \ \sech^2 \left(\sqrt{\frac{E}{2} + b^2N^2} \ t \right )
\end{equation}
The solitonic profile is similar to that of (\ref{nlqm30}) but with a nature for the magnitude of the amplitude and the nature of the argument. \\

Finally, let us remark that an exact solution to equation (\ref{nlqm11}) also exists even when both $b \neq 0$ and $\mu \neq 0$ which is given by the following profile

\begin{equation}\label{nlqm33}
    x (t) = \frac{N^2}{4} \sech^2 \left ( (\mu + b) Nt \right )
\end{equation}
A comparison with the two previous results shows that, as the parameters $\mu$ and $b$ vary, the nature of the amplitude changes, as well as the arguments of the solutions. The distinction is clearly unraveled in Fig.$1$, where as $t$ becomes very large, $x(t)$ makes a gradual drop from its peak value and tends to flatten out. \\

\begin{figure}[ht]
    \centering
    \includegraphics[width=0.75\linewidth]{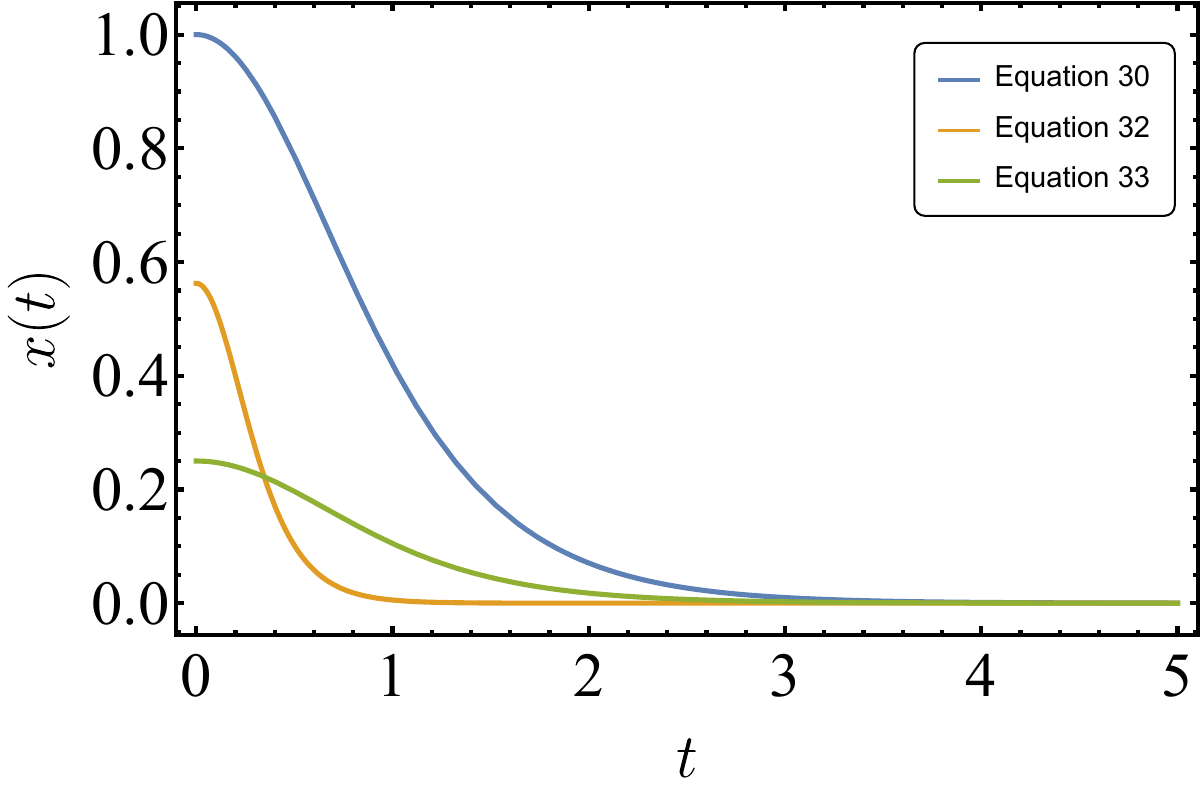}
    \caption{Variation of $x(t)$ vs $t$ for the three equations (\ref{nlqm30}), (\ref{nlqm32}), and (\ref{nlqm33}) for positive values of $t$ only. We have taken $E=10$, $N=1$, $\mu=1$ and $b=-2$ in this plot. }
    \label{fig:placeholder}
\end{figure}

\subsection{Li\'{e}nard type}

One notices that (\ref{nlqm12}) belongs to the Li\'{e}nard class of ODE whose usual dynamical formulation (see, for example, \cite{vill, znojil, bagchi1, ghosh}) in one degree of freedom in the presence of a restoring force and nonlinear damping has the form $ \ddot{u} + f(u) \dot{u} + g(u) = 0$. The coefficients $f(u)$ and $g(u)$ are fundamental in determining the system's symmetry, the existence of periodic orbits, and the bifurcation of limit cycles. In the present case, both the coefficients are odd and are relevant in the context of isochronicity, where all periodic solutions share the same period \cite{bagchi2}. \\

To inquire into other plausible solutions for (\ref{nlqm12}), we pursue the technique proposed in \cite{agc} (see also \cite{poly, nucci}). To this end, we effect the substitution $w(y) = \frac{dy}{dt}$ that facilitates conversion of (\ref{nlqm12}) to the Abel form 

\begin{equation}\label{nlqm16}
w\frac{dw}{dy} = [-2(2\mu + b)y]w -2\mu (\mu + b) y^3 + 2\mu (\mu + b) N^2 y
\end{equation}
Applying a change of variable $w \rightarrow z$ through the transformation $w=yz+ay^2$, converts (\ref{nlqm16}) to

\begin{equation}\label{nlqm17}
(yz+ay^2)\frac{dz}{dy}+ [3a + 2(2\mu+b)]yz+ z^2 +  2a^2y^2= -2(2\mu+b)ay^2-2\mu(\mu+b)y^2+2\mu(\mu+b)N^2
\end{equation}
where note that the substitution $a=-\frac{2}{3}(2\mu+b)$ enables us to get rid of the second term. We are thus left with the first-order form

\begin{equation}\label{nlqm18}
\left[-z^2+2\mu(\mu+b)N^2+\frac{4}{9}(b+\frac{\mu}{2})(b-\mu)y^2\right]\frac{dy}{dz}= \left [yz-\frac{2}{3}(2\mu+b)y^2 \right]
\end{equation}
which yields the Bernoulli form of differential equation for the following two cases of the parameter $b$

\begin{eqnarray}
&& \frac{dy}{dz}=\frac{yz-\mu y^2}{-z^2+\mu^2 N^2},\;\;\;b=-\frac{\mu}{2} \label{nlqm19}\\
&&\frac{dy}{dz}=\frac{yz-2\mu y^2}{-z^2+4\mu^2 N^2},\;\;\;b=\mu \label{nlqm20}
\end{eqnarray}
The explicit solution for $b=-\frac{\mu}{2}$ is given by


\begin{equation}\label{nlqm21}
y (\xi) =\frac{N}{\xi+B|\xi^2-1|^{1/2}}
\end{equation}
where $B$ is an arbitrary constant and we have employed $\xi=z/(\mu N)$. Its profile is illustrated in Fig. 2 for two sample values of $B = 1$ and $2$. These correspond to different input choices of $N$. At $\xi = 1$, a cusp in the solution appears. For $N = 1$, the trajectory shows a definite tilt away from the negative $\xi$-axis in contrast to the positive side, while for both $N=1$ and $N=2$, the profile tends to overlap with both axes as asymptotes. For the other case $b = \mu$, the solution is easily obtained by scaling $\mu \rightarrow 2 \mu$.\\


\begin{figure}[htbp]
    \centering
    \begin{minipage}[b]{0.4\textwidth}
        \centering
        \includegraphics[scale=0.3]{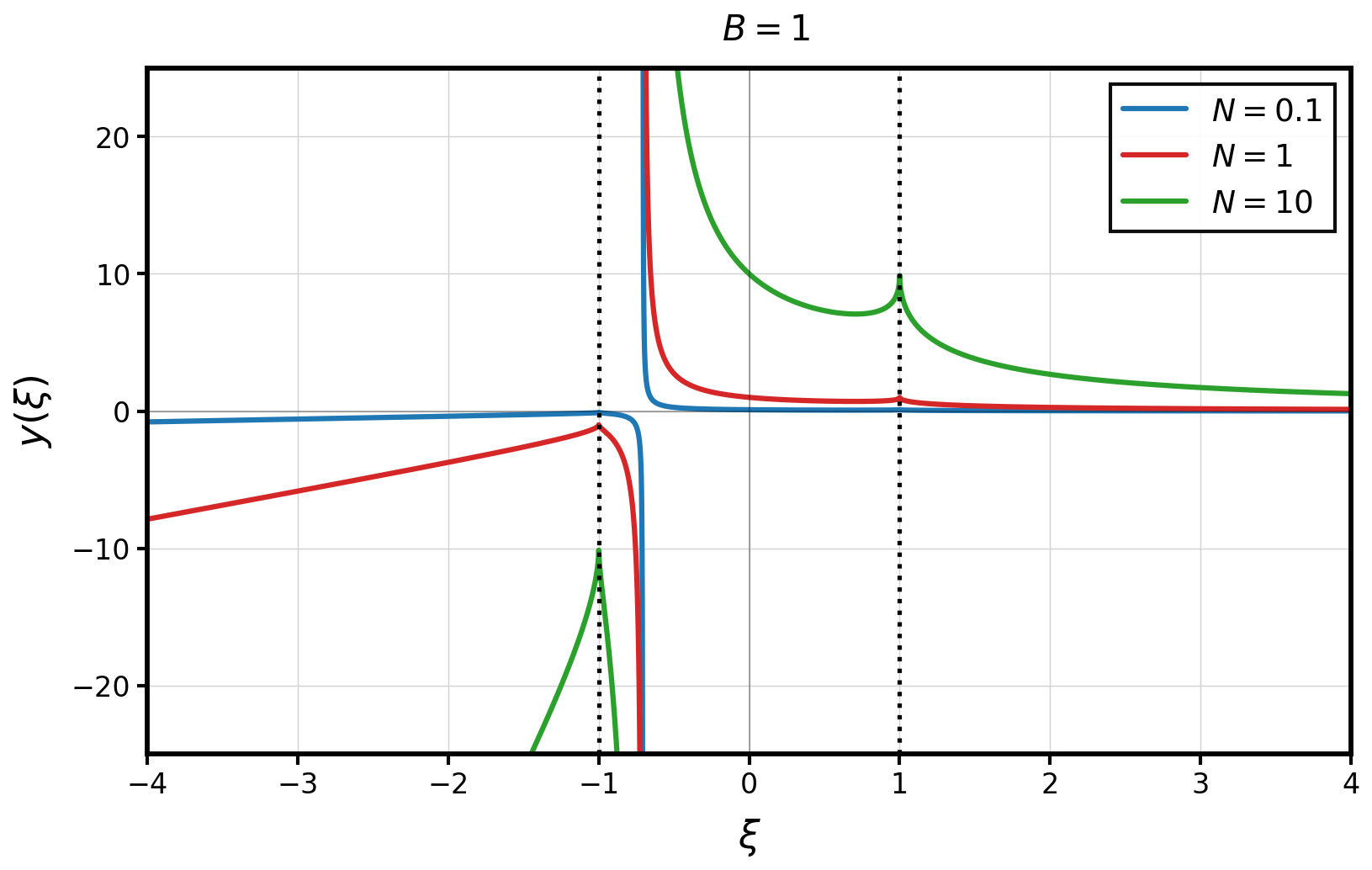}
        \vspace{2pt}
        {\small $(a)$}
    \end{minipage}
    \hspace{10mm}
    \begin{minipage}[b]{0.4\textwidth}
        \centering
        \includegraphics[scale=0.3]{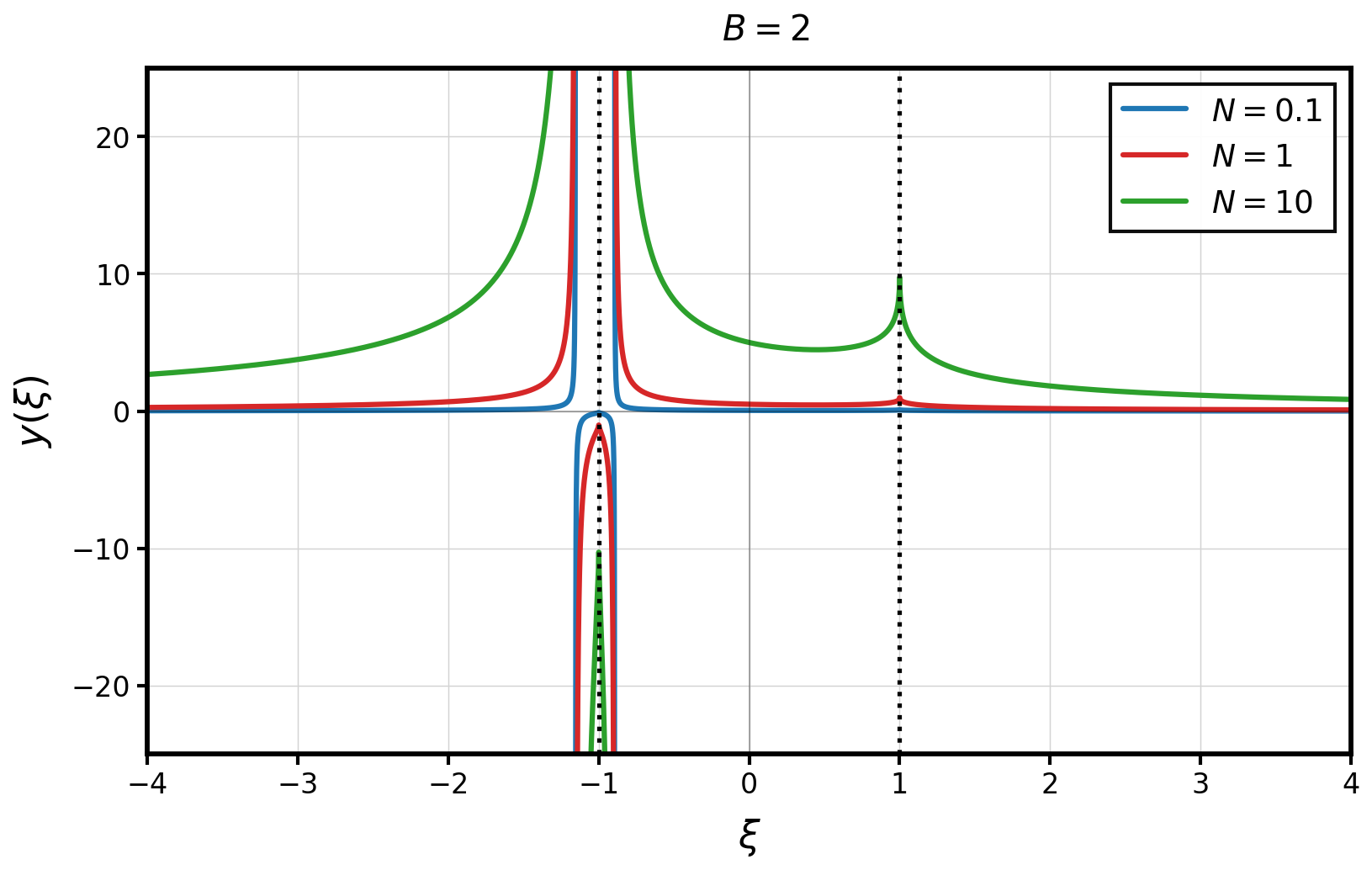}
        \vspace{2pt}
        {\small $(b)$}
    \end{minipage}
    \caption{Plots of $y(\xi)$ vs $\xi$ for values of $B = 1, 2$.}
\end{figure}

\section{Concluding remarks} 

The central question addressed in this paper relates to asking how a nonlinear generalization of quantum mechanics modifies the Schr\"{o}dinger equation, thus acting as a candidate for an effective quantum system where the interaction term accounts for the nonlinear behavior. As already noted in the literature, such generalizations could admit changes resulting from gravity, quantum measurement, or generalized uncertainty principles, apart from other interesting consequences that could follow. For instance, these could relate to the appearence of state-dependent energy spectra resulting from the drastic behavior of quantum states as well as modified wave-packet dynamics, and also from the natural emergence of localized soliton-like solutions, along with their altered stability properties.  In this paper, within a generalized framework of quantum mechanics, we have studied the latter aspect in connection with the recently advanced NLQM scheme of Chodos and Cooper. While these authors attempted at working out simple-minded solutions, our task, apart from obtaining solutions supporting nontrivial equilibrium points that are stable character, has been to go beyond their scenario to demonstrate how Levinson-Smith and Le\'{e}nard equations emerge through the process of disentangling the coupled differential equations, and to explore how new and nontrivial classes of solutions could be generated. The Levinson-Smith equation, which appear frequently in problems of integrable systems, is well known to encompass the Li\'{e}nard family of ODEs which in the present case is typified by coefficients that preserve odd-odd symmetry. The universal utility of Levinson-Smith class of solutions relies on a theorem that specifies strict conditions as to when a unique, stable periodic solution would exist in its phase space. On the other hand, Li\'{e}nard type generalizes damped oscillators and involves coordinate-dependent nonlinear damping force, along with a restoring force term. For the Levinson-Smith class, we showed its relevance to PDM system in quantum mechanics in the presence of either a power-law mass profile or a singular one depending on the choice of the running parameter. We also derived solutions that include a class of solitonic-like profiles by suitably exploiting the condition on the level surface. These solutions indicate the tendency of the energy to remain confined in a localized packet. 
To solve the equation Li\'{e}nard, we followed the procedure of transforming it to Abel's equation and then by applying an appropriate change of variable reduced it to a first-order form that is immediately tractable. Some qualitative features of the solutions have been graphically illustrated and appearence of cusp-like behavior in the solution-profile noted. \\




\textbf{Acknowledgments:} 
The authors thank Sauvik Sen for his generous help in preparing the figures and checking the calculations. \\

\newpage

\textbf{Declarations} \\



\textbf{Funding} 

This research did not receive funding.\\

\textbf{Data-availability statement} 

All data supporting the findings of this study are included in the article.\\






\textbf{Competing interests} 

Both authors declare no conflict of interest.\\

\end{document}